\documentclass{aa}

\usepackage{graphicx}

\begin{document}

\title{Results from a Near Infrared Search for Emission-line Stars in the 
Inner Galaxy: Spectra of New Wolf-Rayet Stars\thanks{Based on observations 
made with ESO Telescopes at the La Silla Observatory under programme ID
P69.D-0567}}

\author{Nicole L. Homeier \inst{1}\inst2
\thanks{Visiting Astronomer, Cerro Tololo 
Inter-American Observatory, National Optical Astronomy Observatories, which 
are operated by the Association of Universities for Research in Astronomy, 
Inc., under cooperative agreement with the National Science Foundation.} 
\and Robert D. Blum \inst{3}$^{\star\star}$ \and Anna Pasquali \inst{4} 
\and Peter S. Conti \inst{5} \and Augusto Damineli \inst{6}}

\offprints{N. Homeier, \email{nhomeier@eso.org}}

\institute{European Southern Observatory, Karl Schwarzschild Str. 2, Garching 
bei Muenchen, Germany \and University of Wisconsin, Madison, Sterling Hall, 
475 N. Charter St., Madison, WI \and
Cerro Tololo Interamerican Observatory, Casilla 603, La Serena, Chile \and 
European Southern Observatory - Space Telescope European Coordinating 
Facility, Garching, Germany \and JILA and APS Department, University of 
Colorado, Boulder, CO 8030 \and Instituto Astronomico \& Geof\'{i}sico, 
S\~{a}o Paulo, Brazil}

\date{Received / Accepted}

\titlerunning{Results from a NIR search for Galactic WRs}
\authorrunning{Homeier et al.}

\abstract{We present follow-up spectroscopy of emission line candidates
detected on near-infrared narrow band images in the inner Galaxy 
(\cite{Hetal03}).
The filters are optimized for the
detection of Wolf-Rayet stars and other objects which exhibit
emission--lines in the 2 $\mu$m region.  Approximately three square
degrees along the Galactic plane have been analyzed in seven
narrow--filters (four emission--lines and three continuum). We have 
discovered 4 new Wolf-Rayet stars and present coordinates, finding charts,
and K-band spectra.}

\maketitle
  

\section{Introduction}

Dust in the plane of our Galaxy severely compromises any attempt at 
complete samples of young stars, but can be mediated by observing at
longer wavelengths. 
In Homeier et al. (2003), we discussed the difficulty in conducting
searches for evolved massive stars in the optical passbands, and detailed our
observing strategy for finding obscured emission-line stars using narrow band
filters near 2~$\mu$m. The eventual goal of such a project is the 
identification of this young stellar population to increase the number of 
known Wolf-Rayet (WR) stars, to improve the statistics of WR stars in 
clusters vs. isolated (likely runaways) WRs, and to use WR stars as a tracer 
of Galactic structure. In this paper we report on spectroscopic follow-up from 
our imaging program in the inner Milky Way.

\section{Observations \& Reduction}

For details of our reduction strategy for the survey images and candidate
selection we refer the reader  
to Homeier et al. (2003). Our candidates are selected with a combination of 
photometric index and image subtraction. We avoided the region within
$5 \arcmin$ of the Galactic center due to severe crowding.

In Table 1 we present central wavelengths and FWHMs for
our chosen set of K-band filters (\cite{BD99}, \cite{Hetal03}).
Four filters are centered on 
the characteristic WR wind emission lines of He I 2.06 $\mu$m, 
C IV 2.08 $\mu$m, H I Br$\gamma$ 2.166 $\mu$m,
and He II 2.189 $\mu$m, and the additional three continuum filters
are at 2.03 $\mu$m, 2.14 $\mu$m, and 2.248 $\mu$m. Thus each line filter
measurement has a continuum point to the red and blue, which 
accounts for spatially variable extinction. 

Spectra were taken with SOFI (\cite{Metal98})
on the European Southern Observatory's (ESO) 
3.6m New Technology Telescope (NTT) 
at La Silla on the nights of 2002 May $1-2, 4-6$ and
with OSIRIS\footnote{
OSIRIS (Ohio State Infrared Imager and Spectrograph) is a collaborative 
project between Ohio State University and CTIO. Osiris was developed 
through NSF grants AST 90-16112 and AST 92-18449. OSIRIS is described in 
the instrument manuals found on the CTIO Web site at http://www.ctio.noao.edu. 
See also DePoy et al. 1993.} 
on the Cerro Tololo Inter-American Observatory's (CTIO) Blanco 4m 
during the nights of 2000 July $12-13$ and 
2002 July $9-12$. Seeing varied between $0.8 \arcsec$
and $2 \arcsec$ during the run at the NTT, between $1-1.5 \arcsec$ 
during the 2000 run at the Blanco 4m, and between $0.8 \arcsec$ and
$1.5 \arcsec$ during the 2002 run at the Blanco 4m.

SOFI and OSIRIS deliver plate scales of $0.29\arcsec$~pixel$^{-1}$ and 
$0.40\arcsec$~pixel$^{-1}$, respectively.
With SOFI we used the GR Grism Red for a wavelength coverage from 
$1.5-2.4 \mu$m, and the the $f/3$ spectroscopic mode on OSIRIS for a 
wavelength coverage of $2.0-2.4 \mu$m.
The spectral resolution at 2.2 m is $\lambda/\delta\lambda\approx 800$
for SOFI and $\approx1200$ for OSIRIS. 

We offset our targets by $5\arcsec$ along the slit and subtracted these
paired observations from each other to remove the background.
Spectra were extracted with the APALL task in IRAF\footnote{IRAF is 
distributed by the National Optical Astronomy Observatories}. A$0-2$V 
stars were observed at a range of airmass, the Br$\gamma$ line was 
removed by linear interpolation, and telluric correction of the 
science spectra was performed with these spectra. 
For the NTT observations, wavelength
calibration was performed using arc spectra. For the 2000 Blanco 
observations, wavelength calibration was accomplished using OH emission
lines (\cite{OO92}). For the 2002 Blanco observations,
rough wavelength calibration was accomplished using telluric features
in the standard star spectrum. Finally, the science spectra were 
normalized to the continuum.

\begin{table}[t]
\caption[]{Filter Description}
\begin{center}
\begin{tabular}{cc}\hline\hline
Central $\lambda$ ($\mu$m) & FWHM ($\mu$m) \\\hline
2.032 & 0.010  \\
2.062 & 0.010  \\
2.077 & 0.015 \\
2.142 & 0.020  \\
2.161 & 0.022 \\
2.191 & 0.013 \\
2.248 & 0.024 \\\hline
\end{tabular}
\end{center}
\end{table}

\section{Results: 4 New Wolf-Rayet Stars}

\begin{figure}
 \center	
 \includegraphics[width=8.5cm,height=8.5cm]{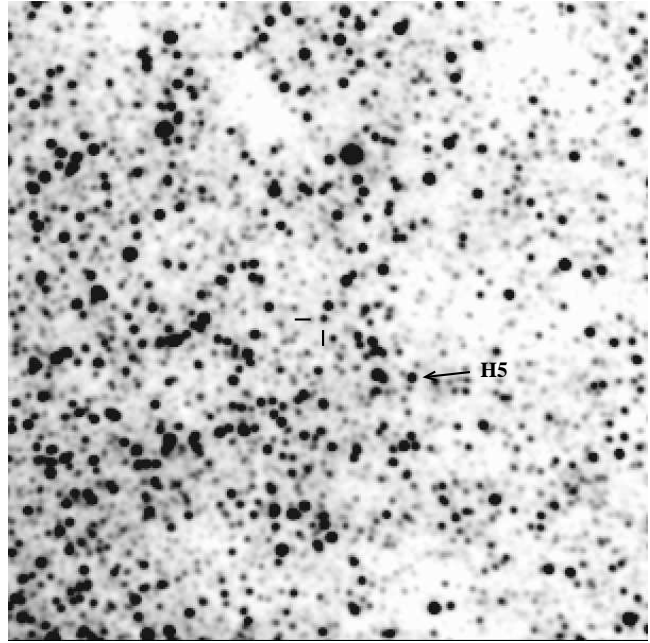}
 \caption{Finding chart for WR101p. The star H5 from Cotera et al. 1999
is $\sim51\arcsec$ to the Southwest.
North is up and east is to the left. The image size is $5 \times 5 \arcmin$.
\label{fig:wr101p}}
\end{figure}

\begin{figure}
 \center	
 \includegraphics[width=8.5cm]{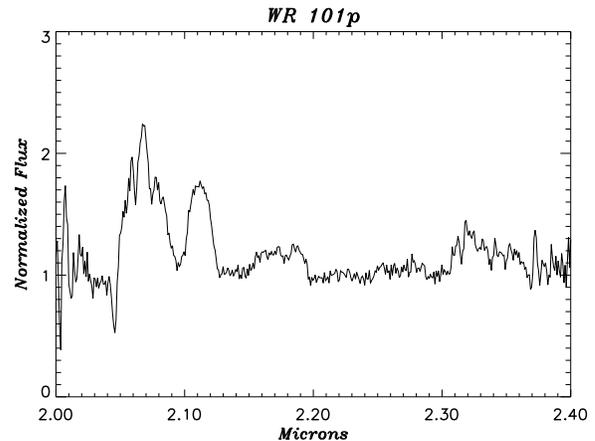}
 \caption{Spectrum of WR101p.
\label{spec:wr101p}}
\end{figure}

\begin{figure}
 \center	
 \includegraphics[width=8.5cm,height=8.5cm]{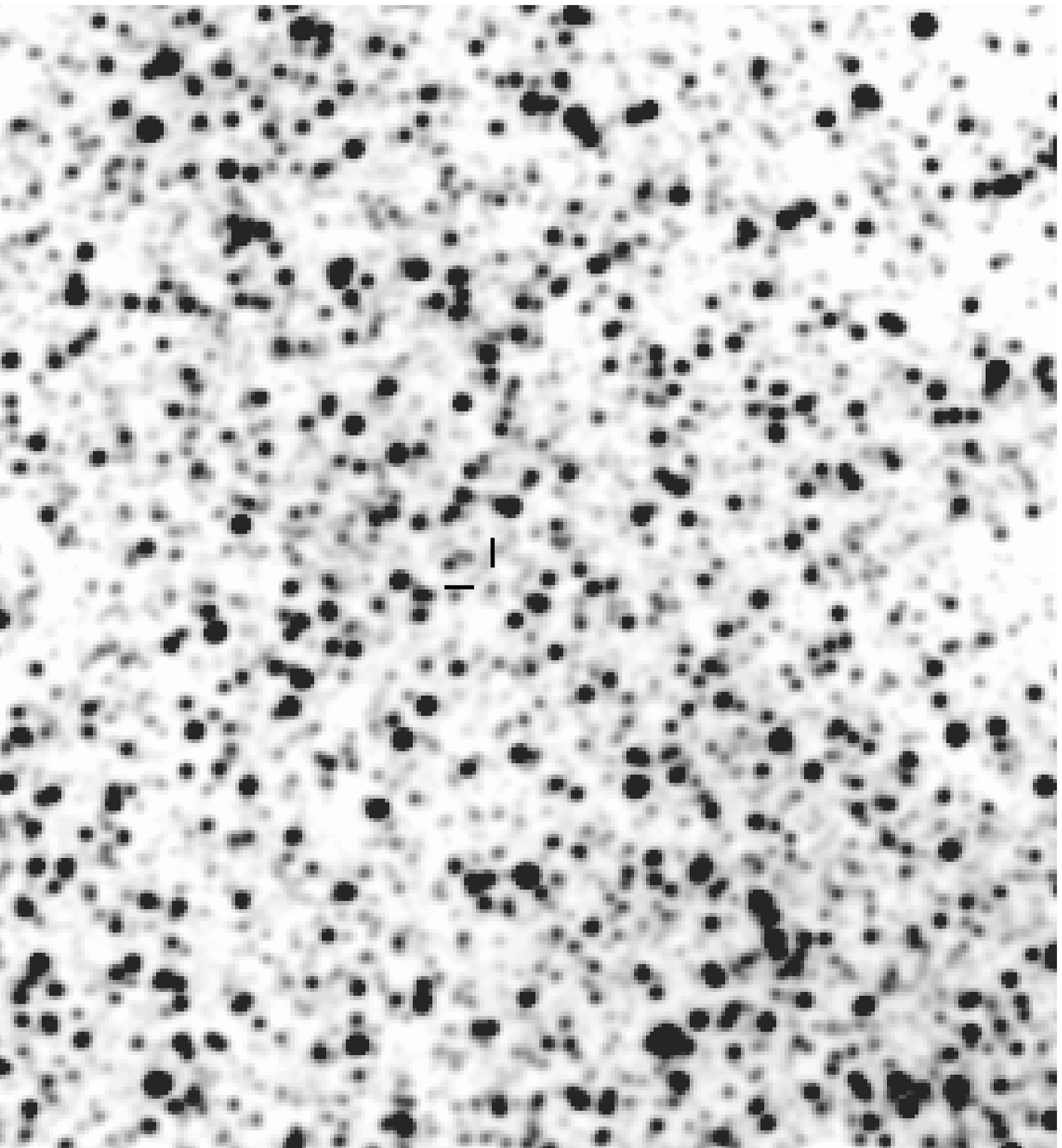}
 \caption{Finding chart for WR101q. 
North is up and east is to the left. The image size is $5 \times 5 \arcmin$.
\label{fig:wr101q}}
\end{figure}

\begin{figure}
 \center	
 \includegraphics[width=8.5cm]{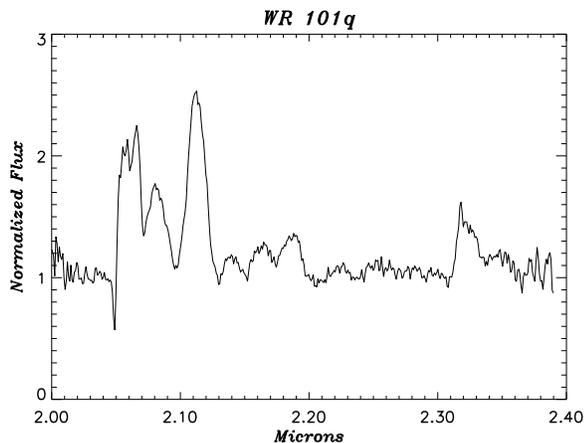}
 \caption{Spectrum of WR101q. 
\label{spec:wr101q}}
\end{figure}

\begin{figure}
 \center	
 \includegraphics[width=8.5cm,height=8.5cm]{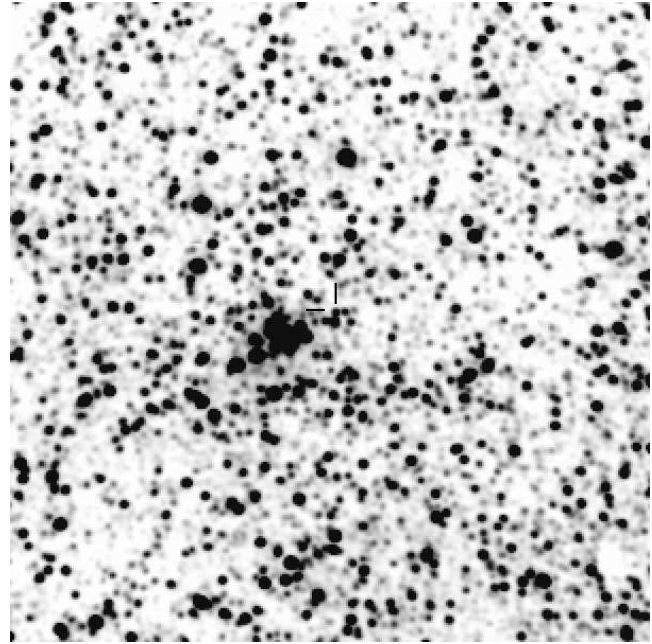}
 \caption{Finding chart for WR102ca. 
The concentration of stars is the Quintuplet cluster, and WR102ca is
a member.
North is up and east is to the left. The image size is $5 \times 5 \arcmin$.
\label{fig:wr102ca}}
\end{figure}

\begin{figure}
 \center	
 \includegraphics[width=8.5cm]{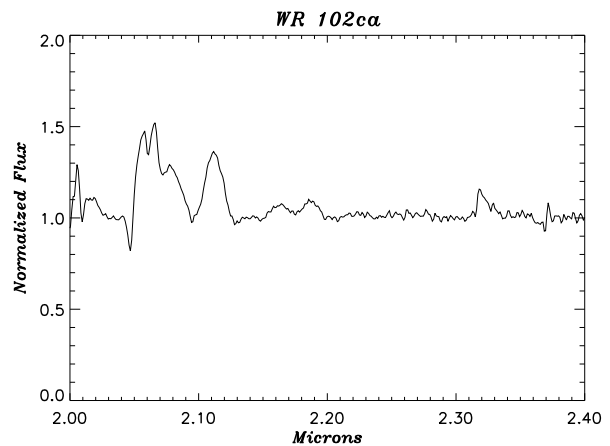}
 \caption{Spectrum of WR102ca. 
\label{spec:wr102ca}}
\end{figure}

\begin{figure}
 \center	
 \includegraphics[width=8.5cm,height=8.5cm]{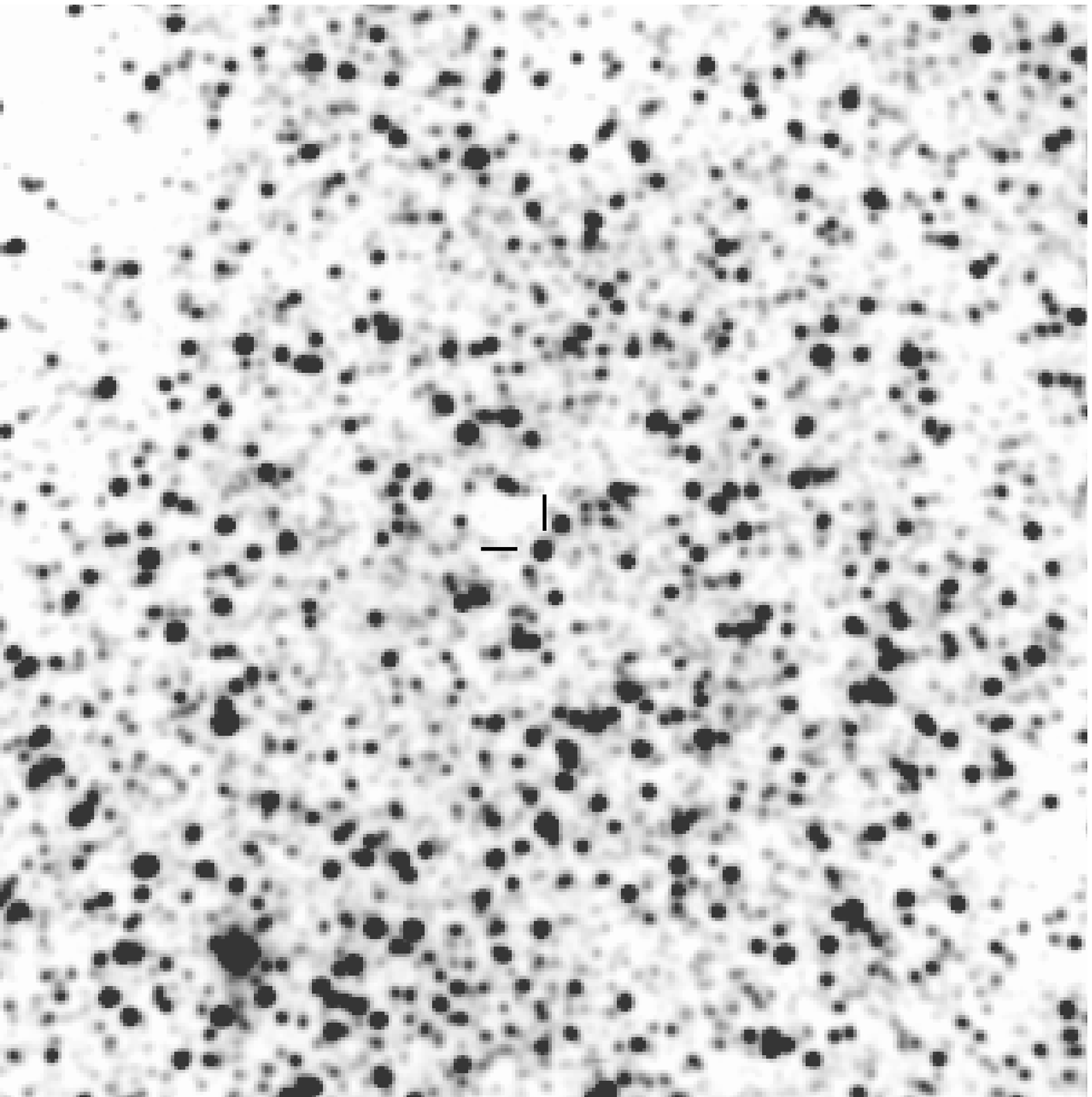}
 \caption{Finding chart for WR102ka. 
North is up and east is to the left. The image size is $5 \times 5 \arcmin$.
\label{fig:wr102ka}}
\end{figure}

\begin{figure}
 \center	
 \includegraphics[width=8.5cm]{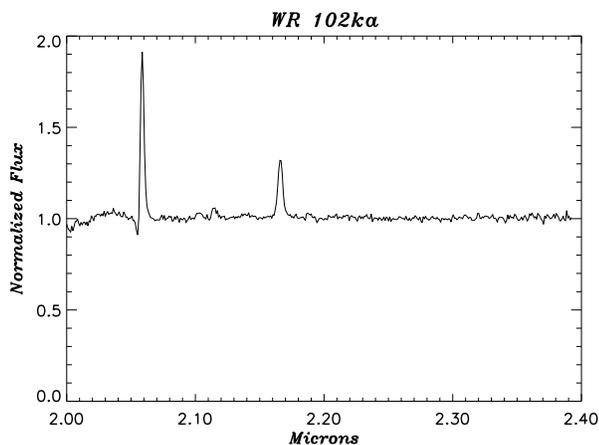}
 \caption{Spectrum of WR102ka. 
\label{spec:wr102ka}}
\end{figure}

We have discovered four Wolf-Rayet stars: three WC and one WN. 
Names were given to the stars following the convention
in the Seventh Catalog of Wolf-Rayet Stars (\cite{vdH01}).
In Table~\ref{jhk} we list coordinates for the four stars and in
Figures \ref{fig:wr101p}-\ref{spec:wr102ka} we present spectra and 
finding charts.

\subsection{Spectra}

\begin{table}
\caption[]{}
\begin{center}
\label{coords}
\begin{tabular}{ccc}\hline\hline
Star & RA (J2000)& Dec (J2000)\\\hline
WR101p & 17:45:42.47 & -28:52:53.3  \\
WR101q & 17:45:57.78 & -28:54:46.1  \\
WR102ca & 17:46:13.04 & -28:49:25.4  \\ 
WR102ka & 17:46:18.12 & -29:01:36.5 \\\hline
\end{tabular}
\end{center}
\end{table}

\begin{figure}
 \center	
 \includegraphics[width=8.5cm]{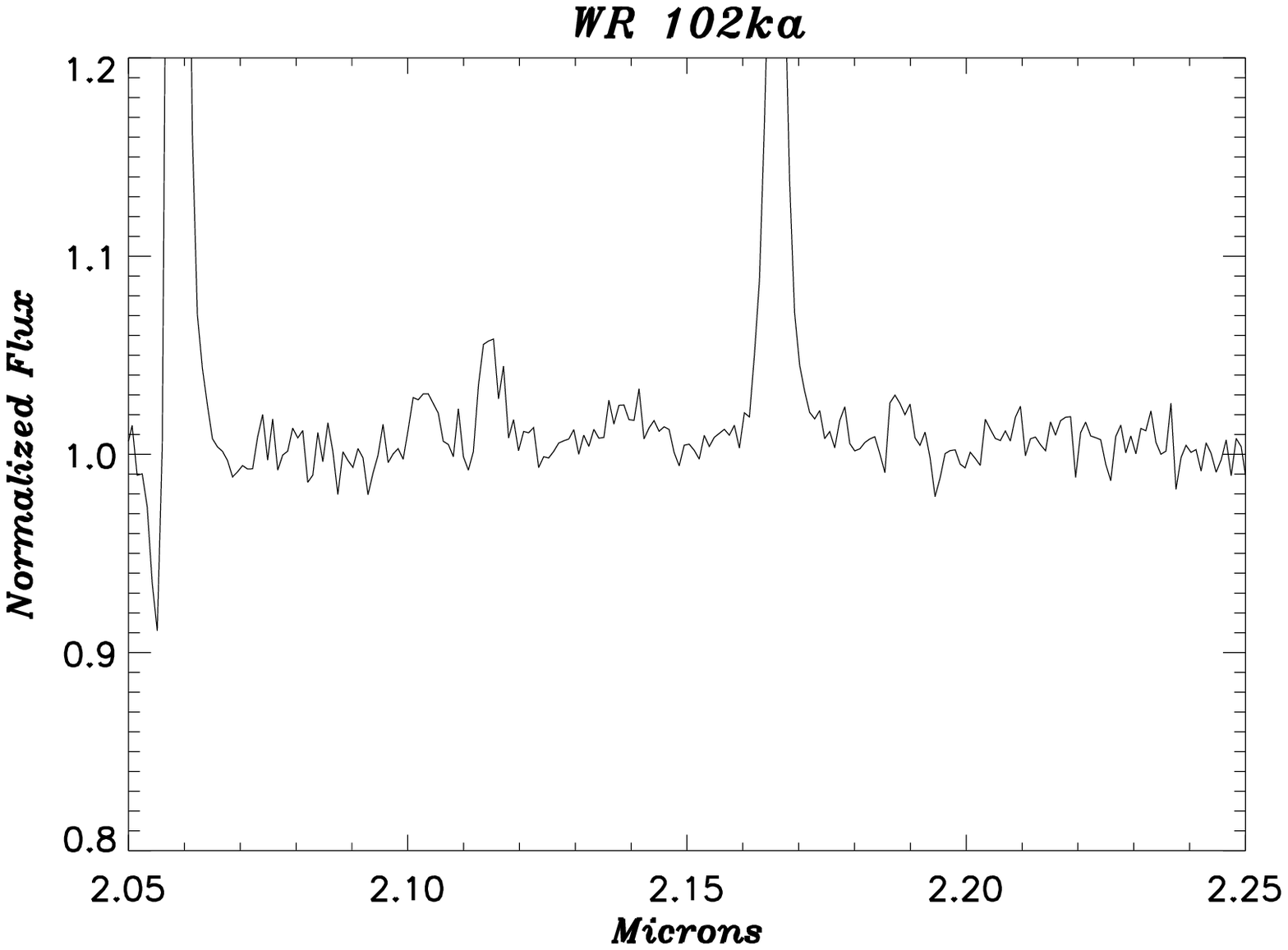}
 \caption{Spectrum of WR102ka on a different scale. 
\label{spec:wr102kazoom}}
\end{figure}

The three WC stars, WR101p, WR101q, and WR102ca, have nearly identical 
$2-2.4 \mu$m spectra. The most significant differences are:  
(1) WR101p lacks the \ion{C}{IV} line at 2.139~$\mu$m, 
(2) WR101q has a \ion{C}{III} line at 
$\sim 2.11 \mu$m which is stronger than the blend of \ion{C}{IV} lines 
at $2.08 \mu$m, in contrast to WR101p and WR102ca,
(3) WR101q has a line at 2.22~$\mu$m of an unknown 
transition, and
(4) WR102ca appears to have more dust than WR101p and WR101q, as its
lines are weaker overall due to excess continuum emission by dust
; see the discussion below on excess emission.

The spectra are similar to the spectrum of star 309 in the 
Quintuplet cluster, which Figer et al. (1999) classified as earlier 
than WC8 due to its excess emission from \ion{He}{II} 3.09~$\mu$m  
and lack of strong emission from \ion{He}{I} 2.058~$\mu$m . 
Based on their similar spectral morphology in the $2-2.4 \mu$m region,
these spectra are consistent with the same classification,
with varying amounts of dust dilution.
Of the published spectra of known WRs other than the Quintuplet WRs, the 
closest morphological match is the WC9 star WR~88 (\cite{EWW91}).
We therefore classify these stars as WC8-9.

The $2-2.4 \mu$m spectrum of WR102ka shows strong lines of \ion{He}{I} 
$2.06 \mu$m and 
Br$\gamma$, weak lines at $2.103 \mu$m \ion{C}{III}/\ion{N}{III}, 
$2.114 \mu$m \ion{He}{I}/\ion{N}{III}, 2.189~$\mu$m \ion{He}{II},
and possibly an unidentified feature at 2.139~$\mu$m. In Figure 
\ref{spec:wr102kazoom} we present the 
$2.05-2.25 \mu$m region of the spectrum of WR102ka to illustrate these
weak features.

Of note is the strength of \ion{He}{I}
$2.06\mu$m to Br$\gamma$ ; the \ion{He}{I}/Br$\gamma$ ratio 
is $\sim 2$. A few of the Galactic Center \ion{He}{I} stars have 
similar \ion{He}{I}/Br$\gamma$ ratios (\cite{Aetal90}, \cite{Ketal91},
\cite{BDS95}, \cite{Tetal96}, \cite{Netal97}), but most also have stronger
$2.11\mu$m features than WR102ka. The most similar spectral morphology 
is displayed by two LMC stars, BAT99~45~(HDE269582, BE294) and 
BAT99~13~(Hen~S9, Sk~$-66~40$) (\cite{Metal88}, \cite{Betal99}).
These stars were originally 
classified as Ofpe/WN9, but later re-classified as WN10h and WN10
(see \cite{CS97} for an explanation). 
The general picture is that stars between $40-60$~M$_{\odot}$ 
evolve through an LBV stage to
become WN8 stars, and in their dormant stage are identified as WN$9-11$ 
stars (\cite{CS97}; \cite{Petal99}).
The star BAT99~45 is at least spectrally variable. The 
HeI/Br$\gamma$ strength changed by a factor of 10 between 1987 January and 
1993 September (\cite{Metal88}, \cite{BDS95}).
Due to the similarity of WR102ka's $2-2.4 \mu$m spectra with BAT99~45
and BAT99~13, we classify it as WN10. 

\subsection{$JHKs$ Photometry}

\begin{figure}
 \center	
 \includegraphics[width=8.5cm]{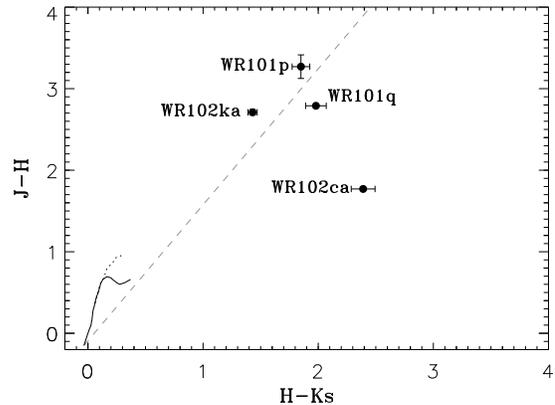}
 \caption{$J-H$ vs.$H-Ks$ plot with magnitudes taken from the 2MASS
catalog; see text. The main sequence and giant tracks are indicated by the 
solid line
and dotted lines (\cite{K83}) and the reddening line is overplotted as the 
dashed line (\cite{RL85}). 
\label{fig:colorcolor}}
\end{figure}

\begin{figure*}
 \center	
 \includegraphics[width=17cm]{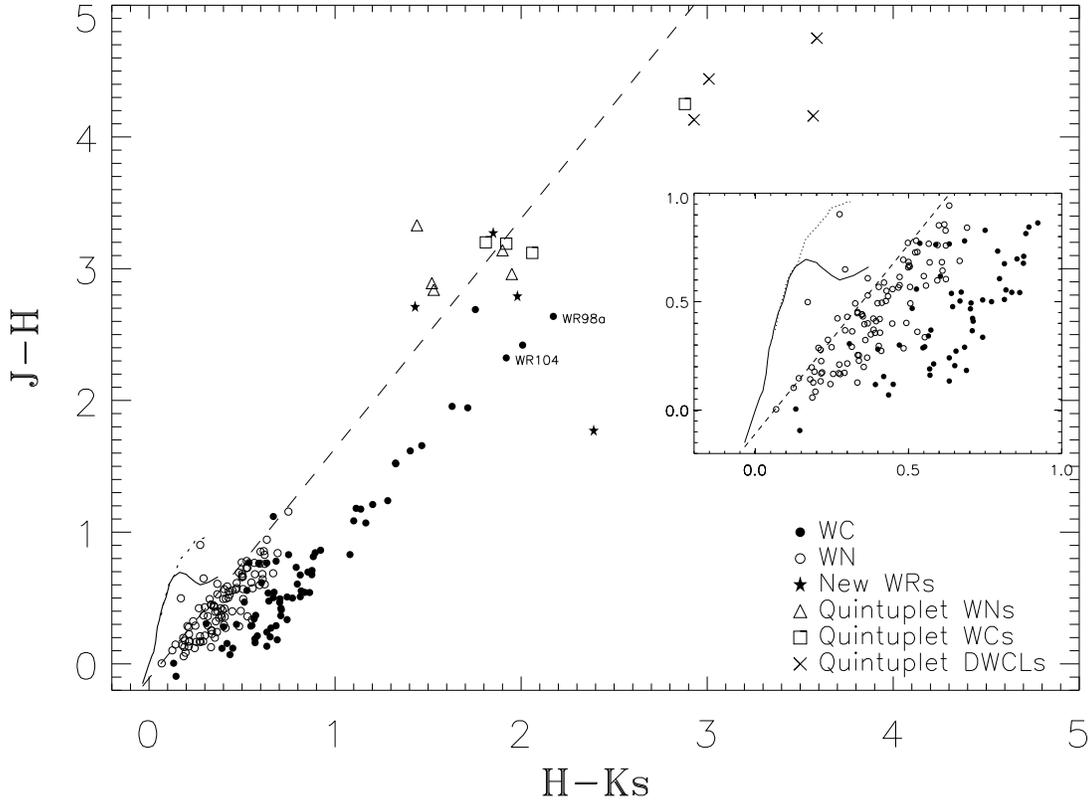}
 \caption{$J-H$ vs.$H-Ks$ plot for a sample of Galactic WN and WC stars, the 
stars discovered here, and the Quintuplet WN, WC, and DWCL stars. The
$J, H, Ks$ magnitudes were retrieved from the 2MASS catalog using the
positions given in the Seventh Catalog of Wolf-Rayet stars (\cite{vdH01}).
Magnitudes are included if the positions in the 2MASS catalog were within
$3\arcsec$ of the input coordinates. If more than one object was found, the 
closest was taken; misidentifications may be possible.
$J, H,$ and K' magnitudes for the Quintuplet stars were taken from 
Figer et al. (1999); where two measurements were made, only the first 
is plotted. The main sequence and giant tracks are indicated by the 
solid line
and dotted lines (\cite{K83}) and the reddening line is overplotted as the 
dashed line (\cite{RL85}).
\label{fig:jhk_all}}
\end{figure*}

\begin{table*}
\caption[]{2MASS Magnitudes}
\label{jhk}
\begin{center}
\begin{tabular}{ccccccccc}\hline
  Name & $J$ & error &  $H$ &  error  &   $Ks$ & error & $J-H$ & $H-Ks$ \\\hline\hline
WR101p	& 16.32 & 0.132 & 13.05 & 0.059 & 11.20 & 0.049 & 3.27 & 1.85 \\
WR101q	& 16.26 & null  & 13.47 & 0.070 & 11.49 & 0.056 & 2.79 & 1.98 \\
WR102ca & 14.56 & null  & 12.79 & 0.071 & 10.40 & 0.075 & 1.77 & 2.39 \\ 
WR102ka	& 12.98 & 0.022 & 10.27 & 0.025 &  8.84 & 0.031 & 2.71 & 1.43 \\	\hline
\end{tabular}
\end{center}
\end{table*}

In Table~\ref{jhk} we present $JHKs$ magnitudes from the latest calibration
of the Two Micron All-Sky Survey (2MASS) catalog\footnote{http://www.ipac.caltech.edu/2mass/} A $J-H$ vs. $H-Ks$ plot is presented in 
Figure~\ref{fig:colorcolor}, with the main sequence and giant tracks as 
solid and dotted lines (\cite{K83}), and the reddening line for the bluest 
stars as a dashed line according to a Rieke \& Lebofsky (1985) extinction 
law. Objects falling to the right 
of the reddening line have colors inconsistent with extinction along the
line of sight for the bluest stars, i.e. they have excess emission at Ks,
usually attributable to hot dust.

$J, H,$ and $Ks$ errors were added in quadrature to produce the error bars in
Figure~\ref{fig:colorcolor}. For WR101q and WR102ca, no J errors were 
listed in the 2MASS catalog, and we have therefore not shown a 
$J-H$ error.

In Figure~\ref{fig:jhk_all} we plot $J-H$ vs. $H-Ks$ values for most of the 
known Wolf-Rayet stars in the Galaxy taken from the 2MASS catalog. The
$J, H, Ks$ magnitudes were retrieved using the
positions given in the Seventh Catalog of Wolf-Rayet Stars (\cite{vdH01}).
Magnitudes are included if the positions in the 2MASS catalog were within
$3\arcsec$ of the input coordinates. If more than one object was found, the 
nearest object was taken; misidentifications are possible.
We also plot $J-H$ and $H-K'$ for the WN and WC stars in
the Quintuplet cluster, and the four stars with featureless $K-$band spectra
referred to as ``cocoon stars'', or ``dusty, late-type WC stars'' 
(DWCLs; \cite{Fetal99}). For reference, the $K$, $Ks$, and $K'$ filters 
have slightly different wavelength coverages: the $Ks$ filter transmission
profile starts $\sim 0.04$~$\mu$m bluer and ends $\sim 0.06$~$\mu$m
bluer than $K$; $K'$ starts $\sim 0.04$~$\mu$m bluer and ends 
$\sim 0.02$~$\mu$m bluer than $Ks$.

The WR stars are offset from the reddening line for the bluest main sequence 
stars towards larger $H-Ks$. To further illustrate this, 
we include an inset of this region in Figure~\ref{fig:jhk_all}. Both the 
WN and WC subtypes follow sequences parallel to the plotted 
reddening line due to intervening interstellar dust, with the WN stars offset 
by $0.1-0.2$ magnitudes and the WC stars by another 0.4 magnitudes.
This can be understood as the effect of the contributions from free-free 
emission in their ionized winds and a blackbody spectrum from hot dust
around most WC stars (see Cohen et al. 1975 and references therein),
and possibly line emission. 
WR stars have strong emission lines in their NIR spectra and this may
affect their broad-band colors. 

We have also labeled 
the positions of WR98a and WR104 in Figure~\ref{fig:jhk_all}, both which are 
associated with 'pinweel'
nebulae (\cite{Tetal99}; \cite{Metal99}). The known pinwheel nebulae arise 
from the interaction of a WC wind with that of a massive companion.
The positions of these stars along
the WC sequence support the idea that the near-infrared colors are driven by
hot dust, and that most WC stars in the Galaxy produce dust (although not all,
as some fall in the region occupied by the WN stars). 

WR101p (WC8-9) and WR102ka (WN10) have near-infrared colors 
more typical of Galactic WN stars and most of the Quintuplet WR stars, 
as does WR101q (WC8-9), although its $J-H$ color
is uncertain due to an unquantified error in its $J$ magnitude.
WR102ca (WC8-9) appears to have a large infrared excess and at the 
same time appears to suffer from less extinction than the rest of the 
Quintuplet WR stars. We note that it is likely that the J magnitude is 
severely in error, which could give it an anomalously blue $J-H$ color and an 
overestimated $K-$band excess. In the 2MASS 
$H$ and $Ks$ images there are two stars within $\sim 0.08 \arcsec$ radius of 
WR102ca, while in the $J$ image only one of the three stars (not WR102ca)
is easily visible. All three stars have listings of ``null'' for their 
estimated J error, and thus their $J$ magnitudes is likely inaccurate.

\section{Null Detections}

In this section we briefly discuss the candidates which turned out not 
to be emission-line stars upon follow-up spectroscopy. These stars
were readily identified as giants by their strong CO absorption lines.
Only the 2.06 and 2.08 $\mu$m line indices were used successfully to 
identify WR stars. It is unclear if this is due to a lack of 
early-type WNs which dominate at the lines of 2.17 and 2.19~$\mu$m, 
a sensitivity issue due to their typically weaker lines,
or a detection bias. Our 2.17 and 2.19 $\mu$m 
line indices had considerably more
scatter than the 2.06 and 2.08 $\mu$m counterparts, leading us to 
conclude that a detection bias is highly likely. We do however note 
that we were able to use the $2.17\mu$m filter images to identify 
several objects with nebular Br$\gamma$ emission.  

During the first set of follow-up observations, objects were selected 
purely on the photometric indices.
Due to the large scatter in these indices, the success rate was low
(3 emission-line objects out of 45 candidates, including 1 WR star).
Other problems included bad pixels and residual images from bright stars.
However, this was largely eliminated with the image subtraction step.
With the inclusion of this step, we were able to identify 
6 emission-line objects, including 3 WR stars, out of 38 candidates. 
We found that selection as a photometric candidate and a
positive identification as a candidate after image subtraction were
separately necessary but not sufficient conditions for the identification of
an emission-line star.

Of the candidates which were not emission-line objects, most were
multiple objects when observed at higher angular resolution while 
positioning the slit. None of the members
of these multiple objects were emission-line stars. All 4 of the identified
WRs were apparently isolated at higher angular resolution. This is a clear
case for the need for observations at higher angular resolution in the 
inner Galaxy where crowding is the dominant source of photometric error. 

\section{Discussion}

\subsection{WC stars 
\label{sect:wcs}}

All three of the WC stars detected have similar late-type WC spectra
(WC8-9) with dilution by dust. WR102ca
is in the Quintuplet cluster, which has a total of 5 members 
(including this new one) with similar spectra. It also contains 4 stars
with featureless K-band spectra, the so-called ``cocoon'' stars,  
(\cite{Fetal99}), which are likely also to be late-type WCs 
with sufficient dust that the continuum emission overwhelms the lines
(\cite{CT01}; \cite{Metal01}).

Dust around WC stars arises in regions of high density, which can be
achieved in the wind-wind interaction zone with a massive companion.
This has been observationally confirmed; there are known periodic 
WC dust-producers where the dust is produced at 
periastron (\cite{Wetal87}; \cite{Vetal98b}), and recently, the 
persistent dust-producers WR104 and WR98a
have been shown to have pinwheel-shaped nebula (\cite{Tetal99}; 
\cite{Metal99}) produced by a short-period ($\sim1$~yr) WR-O binary
system. It is not known if single WC stars can also produce dust
solely through their clumped winds (see \cite{Vetal98a} for a discussion). 
The hint
of copius dusty WCs in the inner Galaxy where the metallicity is 
likely $\ge$~Z$_{\odot}$ may be a clue to required conditions for
dust formation around WC stars. 


\subsection{WRs in the inner Galaxy
\label{sect:innergal}}

In Homeier et al. (2003), an estimate for the number of WRs expected
in the inner Galactic sample was given, following from a simple smooth 
model for
the distribution of stars and gas in the Milky Way. They concluded that
$\sim 28$ WR stars should be detected within the area covered by the imaging
observations in the inner Milky Way. The model used for the estimate
given in Homeier et al. (2003) was normalized to the 
local number density of WR stars assumed, and also depends on the scale 
length of the Galactic disk, and the distribution of WR absolute K 
magnitudes.

A census of WR stars in the region covered by our images (but outside the 
Galactic center) shows that
there are $\sim 35$ known WRs in the Arches cluster (\cite{Netal95}, 
\cite{Cetal96}, \cite{Betal01}, \cite{Fetal02}), 10 in the Quintuplet 
cluster (\cite{Fetal99}), and at 
least 2 apparently isolated WR stars near Sgr A (\cite{Cetal99}), for a 
total of 47. Most of these are brighter than K$=12$, which was the 
limit considered in the estimate. Already we see that the number of known 
WRs exceeds the prediction by this very simple model by nearly a factor of 
two. This suggests a higher star formation rate in the inner Galaxy
compared to the local neighborhood. Near-infrared observations toward the inner
Galaxy have established that this is indeed the case; see for example
the review of the Galactic Center by Morris \& Serabyn (1996).

From this calculation, we might expect that no new WRs would be
found brighter than K$=12$, whereas we have identified 4 in this region.
The addition of these 4 stars is not numerically significant, but 3
of the 4 are notable for their apparently isolated location. 
Either these stars are isolated and are thus runaways, or they are 
the most massive object of an association which has not yet been discovered.
We find no evidence in our images for associations or clusters around these
stars, but given our poor angular resolution, we cannot rule out their
existence. Also, small clusters may not live very long near the Galactic 
center. Numerical simulations predict that small 
($\le 2\times10^{4}$~M$_{\odot}$) clusters 
will disperse sufficiently within a few Myr so that their surface densities
will be indisinguishable from that of their surroundings (\cite{Ketal00}; 
\cite{PZetal02}).

\section{Conclusions}

We have discovered 4 new WR stars in the inner Galaxy: three WC$8-9$ stars 
with similar spectra and one WN10 star. 
One of the WC stars is a member of the Quintuplet 
cluster, while the others are not members of any previously known 
clusters or associations, nor did we find evidence for any
previously unknown clusters or associations. 
We conclude that narrow-band observations at K$-$band wavelengths 
are a potentially efficient way to identify evolved massive stars
in the plane of the Milky Way, and further higher angular resolution work 
is highly desirable.

\acknowledgements{We would like to thank P. Crowther for especially useful
discussions pertaining to WC9 stars and the ambiguity of Ofpe/WN9
K-band spectra. N. H. happily acknowledges the ESO Studentship Programme 
and thanks the University of Wisconsin Graduate School for partial support.
We would like to thank the referee for a careful reading and helpful
comments. This research has made use of the NASA/IPAC Infrared 
Science Archive, which is operated by the Jet Propulsion
Laboratory, California Institute of Technology, under contract with the 
National Aeronautics and Space Administration.}

\end{document}